\documentstyle[preprint,aps]{revtex}

\tightenlines

\newcommand{\lsim}[1]{
\setlength{\unitlength}{12pt}
\begin{picture}(1.4,1.)
\put(.7,-0.3){\makebox(0.0,1.)[t]{$<$}}
\put(.7,-0.3){\makebox(0.0,1.)[b]{$\sim$}}
\end{picture}#1}
\newcommand{\gsim}[2]{
\setlength{\unitlength}{12pt}
\begin{picture}(1.4,1.)
\put(.7,-0.3){\makebox(0.0,1.)[t]{$>$}}
\put(.7,-0.3){\makebox(0.0,1.)[b]{$\sim$}}
\end{picture}#2}
\begin{document}
\draft

\title{Propagation of ultrahigh-energy neutrinos through the Earth}

\author{R. Horvat \\
  ``Rudjer Bo\v skovi\' c'' Institute, P.O.Box 1016, 10001 Zagreb,
Croatia}

\maketitle

\begin{abstract}
The dispersion relation in matter of ultrahigh-energy neutrinos above the
pole of the $W$ resonance ($E_{\nu} \gsim \;\mbox{\rm 10}^{7} 
\;\mbox{\rm GeV} $), is studied. We perform our calculation using the 
real-time formulation of Thermal Field Theory in which the massless limit for 
the $W$ boson is taken. The range of active-to-sterile neutrino oscillation 
parameters for which there is significant mixing enhancement during 
propagation through the interior of the Earth, and therefore significant 
attenuation of neutrino beams in the Earth at high energies, is estimated. 
Finally, this range is considered in view of the cosmological and 
astrophysical constraints. 
\end{abstract}

\newpage

It is now well established \cite{1,2,3} that the Earth's diameter exceeds the
attenuation length of neutrinos with energies greater than $\mbox{\rm
25}\;\mbox{\rm TeV} $. Such an estimate was based on the calculation of the
cross sections for $\nu N $ collisions at ultrahigh energies (UHE), 
($E_{\nu} \gsim
\mbox{\rm 1}\;\mbox{\rm TeV}$). Because of the smallness of the electron mass, $\nu e$ 
interactions are generally considered as negligible with respect to $\nu N $ 
interactions; $\nu N $ interactions therefore provide the dominant signal and 
account for most of the attenuation of neutrino beams in the interior of the 
Earth at ultrahigh energies \cite{1,2}. There is one exception though, the
resonant formation of the intermediate $W^{-}$ boson in $\bar{\nu}e$
interactions in the neighborhood of $E_{\nu}^{res} = M_{W}^{2}/2 m_e \simeq
\mbox{\rm 6.3} \times \mbox{\rm 10}^{15}\;\mbox{\rm eV}$. 

The promising tool for detection of UHE 
cosmic neutrinos by means of neutrino
telescopes \cite{4} consists of recording the long-range muons produced in
charged-current $\nu N $ interactions that occurs in matter surrounding the
detector. Apart from efficient shielding from the flux of atmospheric muons,
such upward-going muon events have the advantage of enhancing the effective
volume in proportion to the range of the produced muons (typically a few
kilometers for $E_{\mu} \simeq \mbox{\rm 10}\;\mbox{\rm TeV}$). For our
purpose it is important to note that the rate for upward-going muons does
not depend only on  the probability for neutrino conversion to a muon with
energy above the threshold energy, but also through the interaction length
which is responsible for the attenuation of the neutrino flux due to
interactions in the Earth's interior. The typical situation that occurs for
$E_{\nu} \lsim \mbox{\rm 10}^{5}\;\mbox{\rm GeV}$ is that the upward rates
depend little \cite{1} on the calculated $\nu N $ cross sections, since the 
enhanced (weakened) interaction rate is nearly compensated by the
enhanced (weakened) attenuation of UHE neutrinos propagating through the
Earth. On the other hand, the detection of cosmic neutrinos at energies of
$\mbox{\rm 10}^{16}\;\mbox{\rm eV}$ or larger is  beset by the problem 
of the increased importance of attenuation of neutrino beams \cite{1,2}. 
However, even so, the upward rates produced by neutrinos from powerful 
radiation sources, like Active Galactic Nuclei (AGN)\cite{5}, should be 
observable in  a detector whose effective area is $A \simeq 0.1\;
\mbox{\rm km}^{2}$.

As for event rates involving electron neutrinos, they are generally smaller
than the muon event rate by the flux ratio (the initial fluxes of UHE
neutrinos originating from AGNs are expected to have a ratio
$\nu_{e}/\nu_{\mu} \simeq 1/2$) times the detector length divided by the
mean muon range, because of the rapid energy loss of electrons (or
annihilation for positrons). Still, it was shown recently \cite{6} that the
Landau-Pomeranchuk-Migdal effect \cite{7} may effectively enhance the
electron range by detecting upward-going air showers initiated by the $\nu_e
$ interaction near the Earth's surface. On the other hand, resonant
$\bar{\nu}e$ scattering contributes significantly to the attenuation of
$\bar{\nu}_{e}$'s, meaning that the flux of electron antineutrinos in the range  
$2 \times  \mbox{\rm 10}^{15}\;\mbox{\rm eV} \leq E_{\nu} \leq 2 \times 
\mbox{\rm 10}^{16} \;\mbox{\rm eV}$ is extinguished for neutrinos traversing
the Earth \cite{1,2}.

In the present paper we are going to consider another mechanism for
attenuation of UHE neutrinos propagating through the Earth, namely, matter
enhanced neutrino oscillations $\nu_e \leftrightarrow \nu_s $, where $s$ is a
sterile neutrino (e.g. a singlet under the gauge symmetry of the Standard
model). We shall be concerned exclusively with the case of $\nu_e
\leftrightarrow \nu_s $ oscillations where the energy of UHE neutrinos
is above the resonant energy, $E_{\nu} \gsim \;  E_{\nu}^{res}$. Owing to the
new form of the effective matter potential in this regime, we are in 
position to  study MSW 
resonance effect \cite{8} previously ignored in the literature. On the other 
hand, the matter effect of the Earth through the standard effective
potentials ($E_{\nu} << E_{\nu}^{res}$), in the region of  oscillation 
parameters relevant for the solar and atmospheric neutrinos as well as those
suggested by the LSND result, is well 
established now \cite{9}. Even more, a new effect of matter-enhanced neutrino 
mixing, based on a maximal constructive interference among transition 
amplitudes, has been discovered recently \cite{10}. In the following, we shall 
first derive the induced mass squared of the electron neutrino with 
$E_{\nu} \gsim \; E_{\nu}^{res}$, then we
find the range of neutrino parameters for which there is matter-enhanced
$\nu_e \leftrightarrow \nu_s $ oscillation during propagation of $\nu_e $
through the Earth and finally we discuss if the established range could
survive constraints from type II supernovae  as well as big bang
nucleosynthesis  on $\nu_e \leftrightarrow \nu_s $ mixing.

Effects of a medium on neutrino propagation is determined by the difference
of potentials, whose standard-model contribution in the context of Thermal
Field Theory (TFT) may  readily be obtained from the relevant thermal
self-energies of a neutrino: charged-current, neutral-current and tadpole.
Of these, only the charged-current and tadpole contribution are relevant
for our consideration. Let us consider the charged-current diagram in more
detail. Since the ``target'' electrons are massive, the $W$ boson may be
considered ``massless'' always when $s \simeq 2E_{\nu}m_e \gg M_{W}^{2}$.
This corresponds to energies in the lab frame $E_{\nu} \gsim \; 6 \times 
\mbox{\rm 10}^{15} \;\mbox{\rm eV}$. In the opposite limit, $s 
\ll M_{W}^{2}$, the $W$ boson should be considered ``massive'' and the usual
contact approximation for the $W$-propagator is adequate. Using the
real-time formulation of TFT, we discover by explicit
calculation that the induced mass squared of $\nu_e $ is equivalent to the
fermion thermal mass squared \cite{11}, 
\begin{equation}
A^{cc}_{\nu } = 
\frac{g^2}{2 {\pi}^{2}} \, \int_{0}^{\infty} k \, dk \, n_{e}(k_0 )
\;\;\;\;\;\;\;\;\;\;\;\;\;\;\; (E_{\nu} \gg E_{\nu}^{res}) \; ,
\end{equation}
where $g \simeq 0.63 $ is the gauge coupling constant. Note that in
contrast to the mass squared induced by the standard MSW potential,
\begin{equation}
A^{cc}_{\nu } = 
2  \sqrt{2}  G_F  N_e E_{\nu} \;\;\;\;\;\;\;\;\;\;\;\;\;\;\; 
(E_{\nu} \ll E_{\nu}^{res}) \; , 
\end{equation}
(1) is independent of neutrino energy and also there is no explicit
dependence on the number density of  electrons. It should be clearly
stated here that actually we are not dealing with field theory in
equilibrium since all the electrons in the medium are bound electrons. Still,
one is allowed to retain the usual real-time formalism by taking the
(11)-component of the electron propagator to be 
\begin{equation}
S_{11}(k) = ( \not{\!k} + m_{e} )
\left( \frac{\textstyle 1}{\textstyle k^{2} - m_{e}^{2} + i \epsilon}
+ 2 \pi i \,
n_{e} (k_{0}) \, \delta (k^{2} - m_{e}^{2} ) \right)\; ,
\end{equation}
where now a bound electron is assigned a distribution $n_{e}(k_0 )$. Thus
(1) describes the plane-wave impulse approximation, which is, for instance,
the basic approximation of electron momentum spectroscopy of atoms and
molecules \cite{12}. 

Although a connection with equilibrium TFT is now established, one may still
wonder why (1) is equivalent to the fermion thermal mass, that is, to the
characteristic mass scale which appears naturally only in the
high-temperature limit of the fermion self-energy. Before going into
details, let us stress that to first order in perturbation theory at high
temperature, the poles of the full fermion propagator are determined just
by the thermal mass \cite{11}. In hot theories, the high-temperature limit
means that the temperature is much larger than 
mass of the particles under consideration ($m_{e}, M_{W}$ in our
case) and the external momenta. Consequently, all particle masses can be
ignored for practical purposes. In the standard electro-weak theory we now
show that the opposite case, when the external momenta are large and the
characteristic scale of the medium is much less than the particle masses,
can faithfully mimic the high-temperature limit. Indeed, when $E_{\nu} \gg
E_{\nu}^{res}$, the four-momentum squared of the $W$ boson is always much
larger than $M_{W}^2 $, and hence the $W$ boson can be considered
``massless''. To get rid of the electron mass, note that it appears in the
numerator of the thermal part of the electron propagator, and both in the
numerator and denominator of the vacuum part of $S_{11}$ [see Eq. (3)]. It is
however trivially to see that $m_e $ from the numerator disappears when
sandwiched between the two electroweak vertices $\gamma_{\mu} L $, where $L
= \frac{1}{2} (1 - \gamma_{5})$ is the projection operator for the
left-chiral fermions. On the other hand, the real part of the neutrino
self-energy is given only by a contribution where a cut is through the
electron line, since  $W$ bosons are absent from the medium. This means
that there is no contribution from the vacuum part of $S_{11}$, where
$m_e $ appears in the denominator . This
completes the proof that the high-energy limit is equivalent to the
high-temperature limit of hot gauge theories. Strictly speaking, since the
thermal mass always involves the thermal effects from both the $W$ boson and
the electron propagators, the only difference is that the former is absent
in (1).  

Using the above interpretation for bound states and keeping
the same normalization as for quasifree states, one can rewrite (1) in the
following form,
\begin{equation}
A^{cc}_{\nu } \simeq 0.2 <k^{-1}> N_e \; ,
\end{equation}  
where $<k>$ is the average momentum of  bound electrons. For the rough
estimates presented here, it is sufficient to assume that $<k^{-1}> \; \simeq 
\; <k>^{-1}$. Let us choose
the average momentum per atom (with the atomic number $Z$) as a quantity of 
interest here. Going back to atomic physics, one can determine this quantity
by applying the Thomas-Fermi method \cite{13} to the calculation of the
total ionization energy of a neutral atom. The result is 
\begin{equation}
<k>^{Z} \simeq 4.6 \; Z^{2/3}\; \mbox{\rm keV}\;.
\end{equation}  
For our purpose, let us recall that the interior of the Earth consists of
two regions of slowly varying density - the core and the mantle, with
particularly strong density change between the lower mantle and outer core.
The density profile of the Earth can be found in \cite{14}. The density of
the mantle increases from $3$ to $5.5 \; \mbox{\rm gcm}^{-3}$ (the average 
value is $4.7 \; \mbox{\rm gcm}^{-3}$ and the average electron fraction is 
$0.49$), 
while the density of the core varies from $10$ to $13 \; \mbox{\rm gcm}^{-3}$ 
(the
average value is $11.8 \; \mbox{\rm gcm}^{-3}$ and the average electron 
fraction is $0.47$). The core comprises heavier elements, presumably nickel 
($Z= 28$). Hence from (5) we have
\begin{equation}
<k>^{Z}_{core} \, \simeq 42 \; \mbox{\rm keV} \;.
\end{equation}
The mantle consists of lighter elements ($Z=8-16$), and our estimate in this
case is 
\begin{equation}
<k>^{Z}_{mantle} \, \simeq 25 \; \mbox{\rm keV} \;.
\end{equation}
Before determination from the resonance condition of a range of neutrino
masses where maximum mixing enhancement may occur, one should consider the
tadpole graph as well. Being a constant independent of the external neutrino
momentum, it is the same as in the standard MSW case 
($E_{\nu} \ll E_{\nu}^{res}$), giving rise to the induced mass squared which 
grows linearly with energy, i.e.,
\begin{equation}
A^{tadpole}_{\nu} = - \sqrt{2}  G_F  N_n E_{\nu} \;,
\end{equation}
where $N_n $ is the neutron number density. Apart from a negative sign in
(8), let us compare the  magnitude of (8) with the charged-current contribution 
(1). It turns out that for $E_{\nu} \gsim \; \mbox{\rm 10}^{9}\;\mbox{\rm GeV}$
(8)
is beginning to dominate over (1). For our purpose, it is however enough to
consider $E_{\nu} \simeq \mbox{\rm 10}^{7}-\mbox{\rm 10}^{8}\;\mbox{\rm
GeV}$ since the short $\nu_e $ interaction length for energies $E_{\nu} >
\mbox{\rm 10}^{9}\;\mbox{\rm GeV}$ means that the flux of electron neutrinos
is extinguished for neutrinos traversing the Earth. Hence for $E_{\nu} \lsim
\;
\mbox{\rm 10}^{9}\;\mbox{\rm GeV}$,  the total induced mass squared for
${\nu}_e $ is essentially given by (1). It is interesting to note that at a
particular energy around $\mbox{\rm 10}^{9}\;\mbox{\rm GeV}$ there is a
nearly complete cancellation of matter effects in the neutrino propagator
due to the sign of (8).    

With the above simplifications and the range of densities in the Earth's
interior as discussed before, one finds from the resonance condition, 
$A^{cc}_{\nu}
\approx \Delta m_{es}^{2}$, a range of neutrino masses where maximal mixing
enhancement may occur,
\begin{equation}
0.07 < \Delta m_{es}^{2}/keV^2 < 0.12 \;\;\;\;\;\;\;\;\;\;\;\;\;\;\;\;\;\;
(10^9 \;GeV \gsim \; E_{\nu} \gsim \; 10^7 \;GeV\;).
\end{equation}  
Here we have taken a small mixing angle, $\cos 2 \theta \approx 1$, in order
to study oscillation enhancement. $\Delta m_{es}^{2} > 0$ in (9) means that
$\nu_s $ is heavier than $\nu_e $. This is also true for $\bar{\nu}_{e} 
\leftrightarrow \bar{\nu}_{s}$ oscillations as $A^{cc}_{\bar{\nu}} > 0 $ (in
contrast to the standard MSW potential where the sign is reversed for
antineutrinos).

Notice that because of the resonance
condition which is energy independent and  the range of densities in the
Earth, the range (9) is actually very small. Even so, it lies in the region
which might be very interesting to astrophysics as well as cosmology. We
recall that the prediction for the sterile neutrino of $\sim $ keV mass is
not in contradiction with any of the present bound. Indeed, the $\sim $ keV
mass is needed if active-to-sterile neutrino oscillations are to solve the
pulsar velocity puzzle \cite{15}. In contrast to active-to-active
oscillations, this solution is not in conflict with the cosmological bound
on stable neutrino masses since the $\sim $ keV mass sterile neutrino has been
proposed as a viable dark-matter candidate \cite{16}.   

It is well known that if we add the reported results from the LSND
collaboration \cite{17} to the list of neutrino anomalies, then the
explanation of all of them requires a four-neutrino scheme with three active
neutrinos $\nu_e $, $\nu_{\mu}$, $\nu_{\tau}$ and one electroweak-singlet
neutrino \cite{18}. This introduces the 'LSND gap' of a few $\mbox{\rm eV}$,
a few orders of magnitude below our preferred range (9). However, the LSND
result is not confirmed (although not completely ruled out) by a similar
KARMEN experiment \cite{19}. 

It is easy to estimate the range of active-to-sterile neutrino parameters
for which there is significant enhancement mixing during propagation through 
the Earth. For significant transitions to developed, it is necessary that the
propagation distance be greater than about a quarter of a wavelength at
resonance \cite{20}. This constraint gives us a lower limit on the mixing angle.
Taking the longest distance through the Earth ($2 R_{earth} $) we have,
\begin{equation}
\frac{\pi E_{\nu}}{\Delta m_{es}^{2} \sin 2 \theta } < 2 R_{earth} \;,
\end{equation}
which for $E_{\nu} = \mbox{\rm 10}^{7}\;\mbox{\rm GeV}$ and $2 R_{earth} =
\mbox{\rm 1.27} \times \mbox{\rm 10}^{9}\;\mbox{\rm cm}$ gives $(\sin 2
\theta )_{core} > \mbox{\rm 2} \times \mbox{\rm 10}^{-3}$ and $(\sin 2
\theta )_{mantle} > \mbox{\rm 3.4} \times \mbox{\rm 10}^{-3}$. 

There is
however a stronger limit on $\sin 2 \theta $ coming from the condition for
unsuppressed oscillations. The oscillation frequency must be real, otherwise
the system is critically overdamped, the oscillations would be fully
incoherent, and hence in fact there will be no oscillations. Since for
neutrino energy in the range $\mbox{\rm 10}^{15}\;\mbox{\rm eV} \leq
E_{\nu} \leq \mbox{\rm 10}^{21} \;\mbox{\rm eV}$, the cross section scales
with $E_{\nu}$ as $\sigma \propto E_{\nu}^{0.4}$ \cite{1,2}, one finds the mean
free path for neutrinos with $E_{\nu} = \mbox{\rm 10}^{7}\;\mbox{\rm GeV}$
to be $l = \mbox{\rm 0.1} \times 2 R_{earth}$. The condition for
unsuppressed oscillations,
\begin{equation}
l_m < 2 \,l \;,
\end{equation}
then gives $(\sin 2 \theta )_{core} > \mbox{\rm 4} \times 
\mbox{\rm 10}^{-2}$ and $(\sin 2 \theta )_{mantle} > \mbox{\rm 7} \times 
\mbox{\rm 10}^{-2}$. Notice that although the Earth is opaque to UHE
neutrinos, the oscillations may proceed unsuppressed whenever the above
requirement is satisfied. 

Let us finally check up the range ($\Delta m_{es}^{2}/keV^2 \simeq \mbox{\rm
0.12}, \sin 2 \theta  > \mbox{\rm 4} \times \mbox{\rm
10}^{-2}$) from a viewpoint of astrophysics and cosmology. The effect of a
resonant $\nu_e \leftrightarrow \nu_s $ mixing on a type II supernovae was
considered in \cite{21}. The bounds on ($\Delta m_{es}^{2}, \sin 2 \theta $)
derived in \cite{21} are valid only if the sterile neutrinos have a mean free 
path larger than the radius of the supernova core after passing the
resonance;
this is the case if $\sin 2 \theta < \mbox{\rm 3} \times \mbox{\rm 10}^{-2}$.
Our range is therefore unaffected by the type II supernovae constraint. On
the other hand, a naive bound on the $\nu_e \leftrightarrow \nu_s $ mixing
from big bang nucleosynthesis was derived, $\Delta m_{es}^{2} \sin^{4}{2 \theta
} \lsim \mbox{\rm 5} \times \mbox{\rm 10}^{-6} \; \mbox{\rm eV}^{2}$
\cite{22}. Notice a disagreement of our preferred range with the above naive
bound. However, the naive calculations ignored the creation of $\nu -
\bar{\nu}$ asymmetries by active-sterile oscillations \cite{23} in the early
universe; these may efficiently suppress ${\nu}_{e} \leftrightarrow 
{\nu}_{s}$ oscillations, and therefore invalidate the conclusions drawn
from naive calculations (even maximal ${\nu}_{\mu}  \leftrightarrow
{\nu}_s $ oscillations as a solution of the atmospheric neutrino anomaly 
\cite{24}  cannot be excluded \cite{25}). The measurement of the flux of UHE 
neutrinos could thus provide us with a new test for this cosmological scenario.  

To summarize, the dispersion relation for electron (anti)neutrinos in the
Earth's interior for energies above the pole of the $W$ resonance, is
derived. Then we have considered MSW oscillations for cosmic neutrinos
traversing the Earth by including the charged-current self energy diagram for
$\nu_e $. We have shown that the range of neutrino masses where maximal
enhancement may occur could be interesting from a viewpoint of astrophysics
and cosmology. Let us finally stress that in order  
to study a nadir angle dependence beyond $\mbox{\rm 34}^{o}$,
where neutrinos always propagate outside the core, a weaker attenuation of a
$\nu_e $ beam would require the inclusion of the tadpole self-energy for
energies beyond $\mbox{\rm 10}^{9}\;\mbox{\rm GeV}$. This interesting case in
now under study.

{\bf Acknowledgments. } The author acknowledges the support of the Croatian
Ministry of Science and
Technology under the contract 1 -- 03 -- 068.

\end{document}